# Optimizing Vapor Cells for Rydberg Atom-Based Electrometer Applications

**Dhriti Maurya,**[1,2*] **Alexandra Artusio-Glimpse,**[3] **Adil Meraki,**[4,5] **Daniel Lopez,**[6] **and Vladimir Aksyuk**[7]

[1] (Associate) Physical Measurement Laboratory, National Institute of Standards and Technology, Gaithersburg, Maryland 20899, USA

[2] Department of Chemistry and Biochemistry, University of Maryland, College Park, MD 20742, USA

[3] Physical Measurement Laboratory, National Institute of Standards and Technology, Boulder, Colorado 20899, USA

[4] (Associate)Physical Measurement Laboratory, National Institute of Standards and Technology, Boulder, Colorado 20899, USA

[5] Physics Department, University of Colorado Boulder, 1234 Main St., Boulder, Colorado 80303, USA

[6] Penn State University NIST Material Research Institute 100 Bureau Dr 325 Broadway State College, PA 16801, USA

[7] Physical Measurement Laboratory, National Institute of Standards and Technology, Gaithersburg, Maryland 20899, USA

*dhriti.maurya@nist.gov

*Abstract*— We present a comprehensive numerical investigation into the radio frequency (RF) field behavior within miniaturized all-glass and hybrid vapor cell geometries designed for Rydberg atom-based electrometry. Using full-vector finite element modeling (FEM), we analyze electromagnetic field enhancement across a wide frequency range (0.05 GHz to 150 GHz) as a function of polarization, incidence angle, and structural configuration. Two primary vapor cell designs are evaluated: translationally invariant "open" cells and periodically structured "supported" cells composed entirely of low-loss glass, as well as hybrid structures incorporating highly doped silicon. Our simulations reveal that the structured all-glass vapor cells exhibit sharp, angle and polarization-dependent resonant peaks due to guided-mode coupling, resulting in localized RF power enhancements exceeding 8x. In contrast, silicon-based structures demonstrate significant electric field attenuation and suppression of resonant features due to their high dielectric losses. Through k-vector and angle-resolved analyses, we show how cell geometry and material properties critically influence the RF field distribution and coupling efficiency. Our findings open new possibilities for optimizing vapor cell architectures to enhance field sensitivity, directional and polarization selectivity, and integration potential in chip-scale quantum sensing platforms based on Rydberg atoms.

I. INTRODUCTION

Rydberg atom electrometry is a cutting-edge technique that utilizes the extreme sensitivity of Rydberg atoms to electric fields, making it highly effective for precise field measurements [1-4]. Alkali atoms, when excited to high principal quantum numbers, experience large electromagnetic field coupling, making them ideal for accurate electric field detection across a wide and tunable frequency range [5]. A key component in such measurements is the vapor cell, which houses the Rydberg atoms [6-9]. Traditional vapor cells, produced by glass blowing techniques, are large glass tubes, typically ranging from 25 mm in diameter and 25 mm to 100 mm in length. Such cells are difficult to miniaturize below about 5 mm due to the high thermal stress in glass during the forming process, which leads to non-uniform shapes and variations in the glass permittivity and morphology. Development of miniature atomic cells with planar geometries that can be mass produced by batch fabrication and tightly integrated with photonic integrated circuits (PIC) is of high interest for realizing the next generation of atomic vapor-based quantum sensors and systems. Micro-Electro-Mechanical Systems (MEMS) microfabrication has been an enabling technology that combines advantages of batch fabrication with lithographical control of the cell size and structure [10-11]. Notably, in Rydberg atom electrometry the interaction of the electric fields with the specific structure and the materials of the vapor cell presents both a challenge and an engineering opportunity to tailor and harness the interaction [12-15]. Depending on the interplay between cell geometry and its constituent materials' dielectric permittivities and losses at the specific electromagnetic frequencies of interest, interaction effects may include absorption, scattering and near-field enhancement as well as localized and traveling-wave radiofrequency (RF) resonances with sharp incidence angle and frequency dependencies.

Our group has recently developed techniques for wafer-scale fabrication of all-glass millimeter scale planar Cs vapor cells and demonstrated their suitability for Rydberg electrometry with narrow linewidths and stability over extended time [16]. In the long term these microfabrication techniques open the field to a much broader selection of cell materials with different dielectric properties and enable a wide range of accurately fabricated cell geometries with millimeter and sub-millimeter scale features. This motivates the development of better understanding of the interplay between the RF field and the cell geometry and material properties to design enhanced sensors.

In this work, we numerically investigate the millimeter-wave electromagnetic (EM) field distributions in either all-glass or traditional glass-Si MEMS vapor cells for Rydberg electrometry, which have different millimeter-scale geometries. Full-vector finite element method (FEM) EM scattering simulations are performed to characterize the electric fields inside the atom-filled volumes as a function of frequency, angle of incidence, and polarization of an incident EM plane wave. The analysis focuses on three representative geometries: periodically structured all-glass cells, translationally invariant two-glass-windows cells, and periodic cells incorporating doped, conductive silicon middle layers. For numerical efficiency each configuration is modeled in 2D as a single 2 mm periodic unit with Floquet (periodic with a fixed, user-defined phase shift) boundary conditions, neglecting the finite overall size effects. The field enhancements relative to the incident for the frequencies between 50 MHz and 150 GHz and incident angles between surface-normal, 0°, and grazing, 80°, are obtained. The data reveal local field enhancements as high as 2.9x due to the coupling of the incident plane waves into broad standing- and sharp traveling-wave resonances in the open and structured dielectric cells. At other angles and frequencies, we observe broad field suppression due to reflection and absorption losses in the cells with Si. In the structured cells, plotting the enhancement as a function of frequency and incident wavevector reveals incident EM resonant coupling to the quasi-guided cell modes. Our modeling offers insight into the electromagnetic scattering effects in these microstructures under realistic operating conditions, pointing the way to application-specific vapor cell engineering for millimeter-wave electrometry.

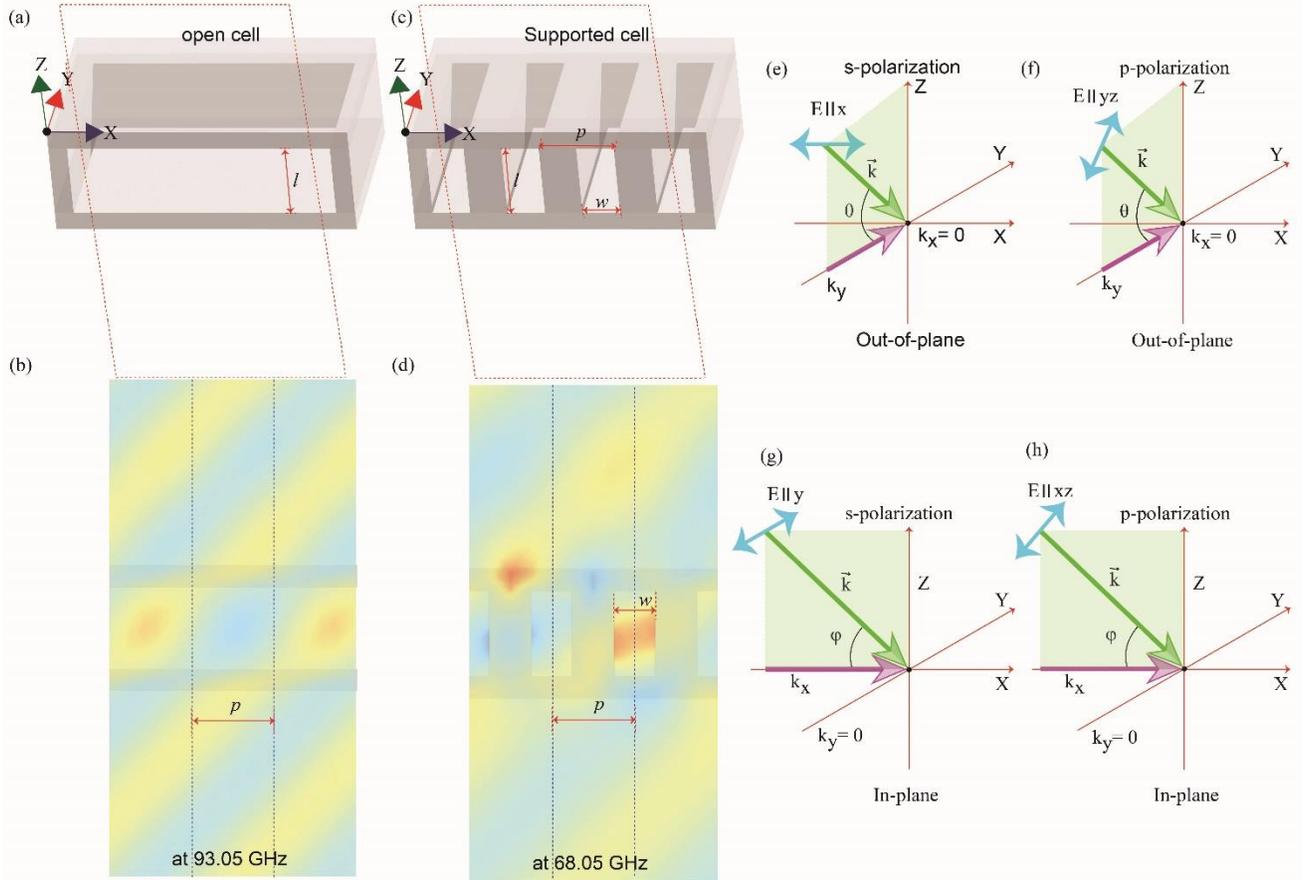

Fig. 1: Schematic of a millimeter-scale Rydberg sensor array incorporating glass vapor cells: 3D representation of the vapor cell design, including (a) an open and (c) a supported vapor cell. Finite element modeling was employed to analyze the RF field distribution and average field enhancement under plane-wave RF incidence within the open and supported cells approximated as infinite translationally-invariant in Y and infinite periodic in X, allowing for 2D modeling in the XZ plane. Modeled cell parameters: top/bottom glass = 0.5 mm, vertical gap $l$ = 2 mm; supported cell (c) period $p$ = 2 mm and horizontal gap $w$ = 1 mm(b) and (d) show the x-component of the RF electric field within the XZ plane inside the open and supported cells, respectively, calculated for P-polarized 46° incidence in the XZ plane at frequencies of 93.05 GHz and 68.05 GHz corresponding to local maxima of field-enhancement. The dotted lines denote simulation domain left and right boundaries with Floquet-periodic conditions, while for illustration the image is extended to three periods using Floquet periodicity. Panels (e-h) illustrate the four specific incidence direction and polarization configurations utilized throughout this study.

## II. VAPOR CELL SCHEMATIC AND DISCUSSIONS

The modeled physical vapor cell geometries, shown in Fig.1 (a & c), correspond to the all-glass extended-area planar microfabricated cells [16] recently developed for millimeter-wave electrometers. These cells extend >15 mm in X and Y to enable spatially resolved field measurements, as either an open cell with a single large cavity between parallel windows supported at the cell edges, or a supported cell featuring multiple parallel channels separated by periodic glass supports designed to accurately maintain optical planarity of the cell windows against the pressure difference between the near-vacuum inside and the ambient. The modeled cells have the top and bottom glass layers measuring 0.5 mm each, and a vertical gap $l$ = 2 mm, set by the middle layer thickness. The supported cell in Fig. 1c has a period $p$ = 2 mm, the horizontal gap $w$ = 1 mm, with a wall width $p - w$ = 1 mm, approximating the dimensions found in the experimental cells. For consistency, the open cell is modeled as periodic with the same period $p$. The Borofloat 33 glass is modeled as a low-loss dielectric with a frequency-independent $\varepsilon$ = 4.481+i0.0817 over the relevant RF frequency range [17]. The atom-filled volumes are assumed to have a dielectric constant of 1. To explore material dependence, a comparison is made with the supported cell of the same geometry with Borofloat 33 windows bonded to a doped silicon middle layer, modeled using a dielectric constant $\varepsilon = 11.7$ and resistivity of 0.1 $\Omega\cdot$m.

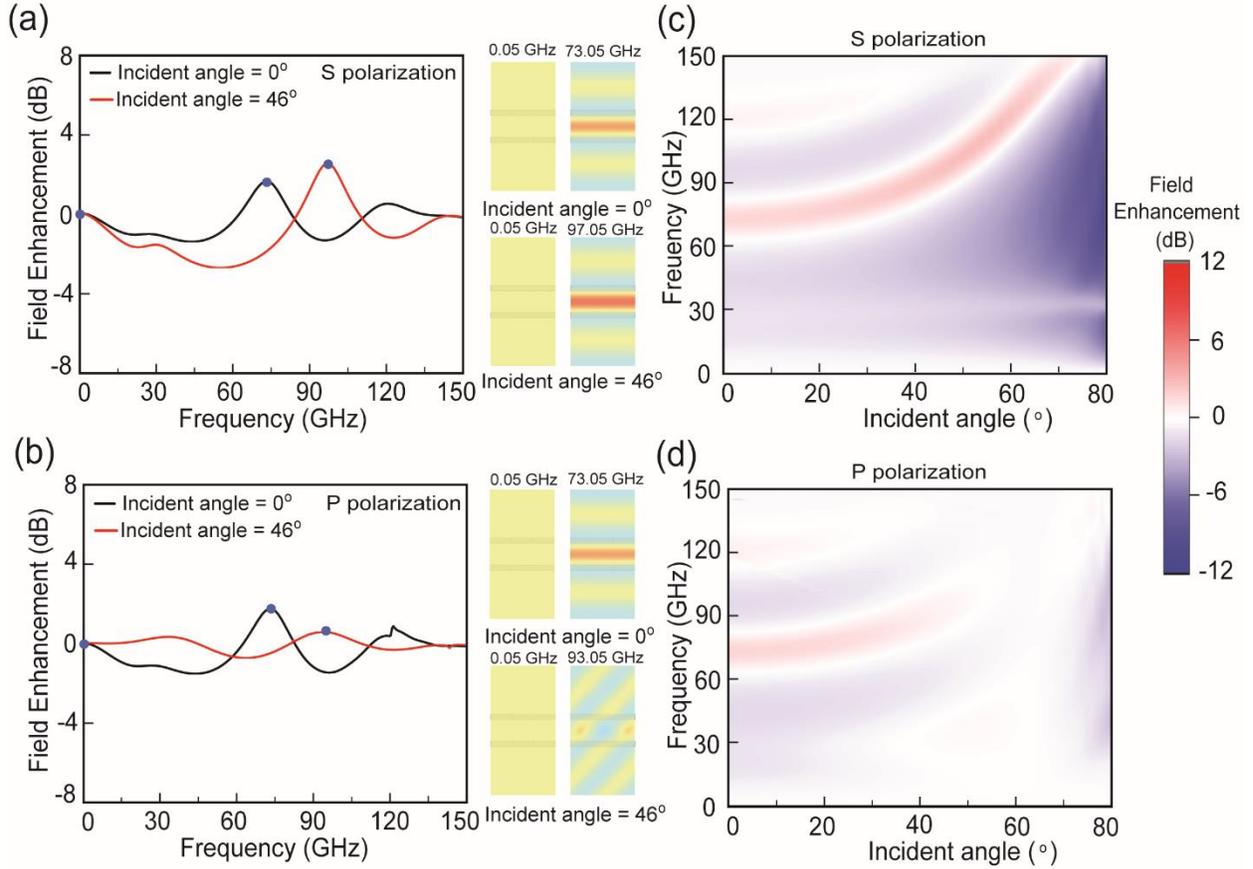

Fig. 2: Numerically calculated time and spatially averaged RF field enhancement relative to the incident field within a glass open vapor cell for S-polarized (a, c) and P-polarized (b, d) incident plane wave as a function of angle and frequency. The out-of-plane incidence is used for S-polarized waves, while the in-plane incidence is used for P-polarized waves, such that the field is in the yz-plane. Panels (a) and (b) show frequency-dependent enhancement for two specific angles, corresponding to specific vertical slices through enhancement color maps presented in panels (c), (d) respectively. The right panels in (a) and (b) show the spatial dependence of the x-component of the RF electric field ($E_x$) within the yz-plane of the 2D domain, evaluated at low frequency (0.05 GHz) and at the peak enhancement frequencies, marked by blue circles. Cell walls are overlayed in semi-transparent gray. All electric field profiles and all field enhancement maps are shown using the same color scales throughout the paper to facilitate direct comparison between different types of cells.

These millimeter-scale cell structures are comparable to the RF wavelength, affecting the local RF fields in the cell. The wavelength-scale periodicity and low loss in the supported glass cell can result in particularly sharp frequency- and angle-selective resonant enhancement of the incident field. Modeling of these effects is crucial for assessing application limitations and potential benefits, such as engineered field sensitivity enhancement and direction, polarization, and frequency selectivity. We employ Finite Element Method (FEM) simulations to solve for the full vector E field distribution inside the cells at specific frequency and incident wave parameters [18]. Simplifying the cell as infinitely extended, translationally invariant in Y and infinitely periodic in X and specifying fixed incident wavevector components $k_x$, $k_y$ reduces the problem to two dimensions (2D). A commercial FEM numerical solver is used, with the 2D computational domain in the xz-plane representing a single unit cell with a period $p = 2$ mm, surrounded by Floquet-periodic boundary conditions, marked by black dashed lines in Fig. 1(b) and 1(d). A series of full-vector solutions are found for incident RF plane waves with frequencies from 0.05 GHz to 150 GHz equally spaced at 0.5 GHz and angles from 0° to 80° from normal (z) with 2° increments. The in-plane angle of incidence ($\varphi$) in the XZ plane of the domain is defined in the model by setting the wave vector component $k_x$ and the out-of-plane angle ($\theta$) is defined by $k_y$. The modeling domain includes 12 mm of empty space above and below the cell's top and bottom window surfaces, terminating at port boundaries parallel to the cell, backed by Perfectly Matched Layers (PMLs) vertically extending for 20 mm above the top and below the bottom ports. Incident RF plane wave is specified at the top port. Examples of the resulting solutions, with the cell structure overlay in semi-transparent gray, are shown in Fig.1(b &d), extended laterally for three periods using Floquet periodicity.

While the setup allows modeling plane wave scattering from any incident direction and with any polarization by appropriately specifying $k_x$, $k_y$, and the incident port fields, we model the electromagnetic fields for two distinct incident directions (in-plane and out-of-plane relative to the xz domain) with two linear polarizations (s and p) for each direction, as illustrated in Fig. 1(e-h). Since the intended application involves optically interrogating atomic responses to the magnitude of the electric field using a narrow laser beam, we report the averaged field enhancement factor along the laser path. This factor is defined as the ratio of the E-field magnitude averaged inside the cell along the vertical line through the atomic volume (in-cavity section of the cell boundary, dashed line in Fig.1(b &d)) to the incident field magnitude. In the supported cell case, this boundary line runs through the middle of the sensing volume, which would be filled with Rydberg atoms. The line averaging is used to represent narrow interrogating laser beams. Averaging over the full cell volume produces qualitatively similar results with minor quantitative differences. We also note that the cell generally changes the local electric field polarization state for P polarized incidence, which is important for the quantitative account of the laser Rydberg sensing. Our model can be used to quantify this polarization dependence, but detailing it is beyond the scope of this work.

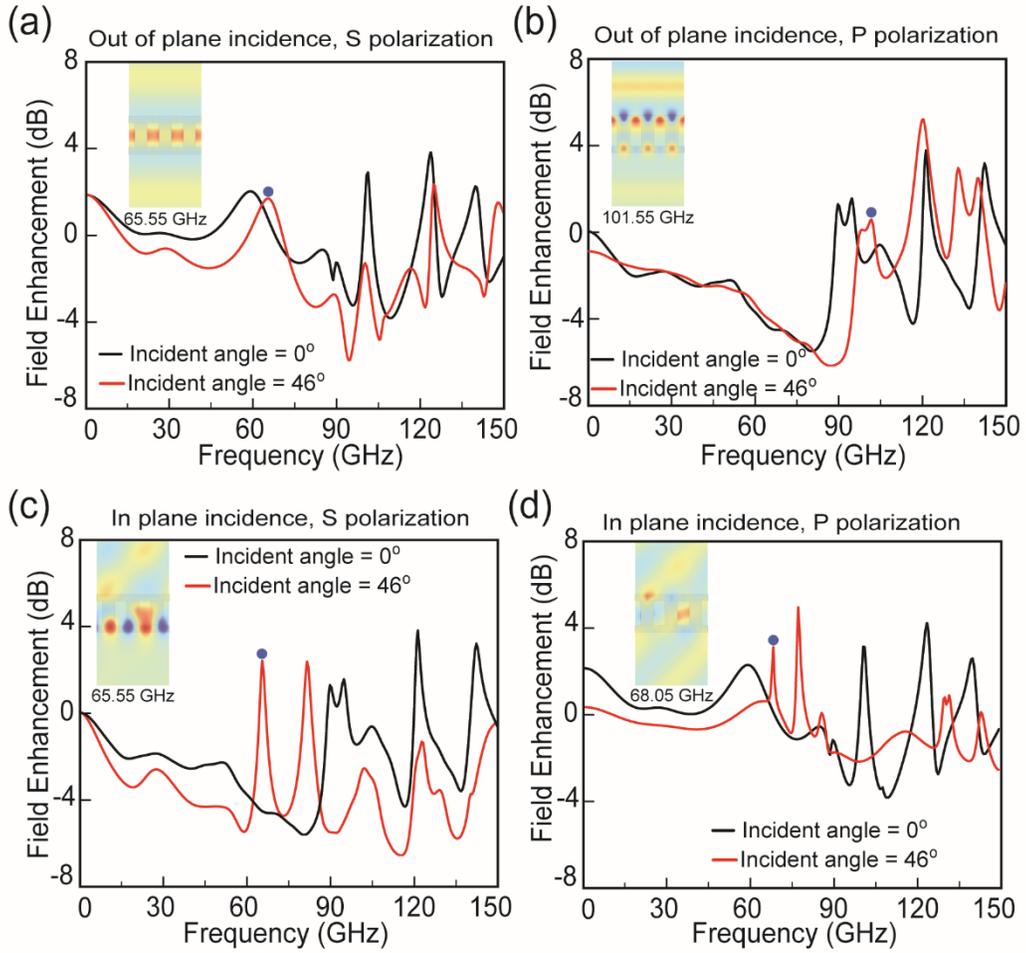

Fig. 3: Numerically calculated RF field enhancement within a supported all-glass vapor cell for two incident angles, 0° and 46°, represented by the black and red solid lines, respectively, under four distinct excitation conditions: (a) out-of-plane incidence with S polarization, (b) out-of-plane incidence with P polarization, (c) in-plane incidence with S polarization, and (d) in-plane incidence with P polarization. Insets in each panel show the corresponding electric field distribution at the peak of field enhancement (indicated by the blue marker) for an incidence angle of 46°. The inset in (a) and (d) displays the $E_x$-component of the RF electric field vector, while the insets in (b) and (c) show $(E_y + E_z)/\sqrt{2}$ within the yz-plane of the 2D domain inside the supported all-glass vapor cell. All incident-normalized E-field profiles are presented on the same scale and show the real component of the field.

Fig. 2 shows the field enhancement results for an open vapor cell for the two different polarizations across frequencies and angles of incidence. Among the four FEM model setups (Fig. 1, e-h), results were substantially identical for the models with the same polarization, irrespective of the incidence direction, as expected for the open cell, which is rotationally symmetric around the z axis (see supplemental Fig. S1, for comparison). The results in the limit of normal incidence are also identical for two polarizations, as expected (black lines in Fig. 2 (a & b), with minor differences attributed to numerical errors. The field enhancement tends to unity (0 dB) in the electrostatic limit of low frequencies irrespective of the angle or polarization. For each polarization, within the broad range of angles we observe a single broad but well-defined enhancement peak as a function of frequency, with some additional broad features. This peak corresponds to the fundamental resonance of the cavity formed by the plane-parallel dielectric windows partially reflecting the RF field. This resonance appears as an increase in the E field inside the cell visible in the electric field maps on the right-hand side in Fig. 2(a & b). The upward frequency shift of the enhancement peak with increased angle is due to the decrease in $k_z$ with angle, which is being compensated by the increase in frequency. Finally, substantial field suppression is notable for the S polarization at grazing incidence angles (blue region in Fig. 2 c), which closely matches with decreased RF transmission and increased reflectivity of the cell in the far field (supplemental Fig. S2).

Next, we investigated the field enhancement in the supported cell (Fig.1(b)) for four different polarization states (shown in Fig. 1(e–h)), with the results presented in Fig. 3 and Fig. 4. At frequencies below 60 GHz no sharp enhancement features are observed. A broad enhancement peak is evident near 60 GHz for the polarization with electric field in the XZ plane (Fig. 3 (a & b) and Fig. 4(a & d) and Fig. 3 a, inset) which is the standing wave resonance similar to the one in the open cell. In the low frequency, electrostatic limit unity enhancement is observed for the cases with the field parallel to the Y axis, i.e. having no components normal to any of the cell walls (Fig. 3c, 4c at all angles and Fig. 3b, 4b at 0°). In contrast, in cases with field normal to the Y axis and the glass ridges, in the low frequency limit the field is electrostatically enhanced in the cell volumes outside the polarizable glass ridges (Fig. 3a, 4a at all angles and Fig. 3d and 4d particularly at small angles where the ridge-normal Ex component is largest). In contrast to the open cell, where translational symmetry forbids coupling of the incident field to the guided RF modes localized to the dielectric structure, scattering on the periodic grating-like structure of the supported cell allows phase matching needed for the excitation of these guided modes at specific frequencies where the frequency- and phase-matching conditions are met. Such grating resonances, well known in photonics, involve energy buildup in the guided RF modes tightly bound to the cell (e.g. Fig. 3 b, c & d insets), manifesting as a series of sharp field enhancement peaks (Fig. 3) occurring above 60 GHz and smoothly varying with incidence angle (Fig. 4) to satisfy phase and frequency matching to the dispersion relationships of various guided modes. The strongest field enhancement of ≈2.9x is observed at 114.5 GHz with in-plane incidence at 12° at P polarization, marked in Fig 4.

We note that at normal incidence (0° angle), the field enhancement results in Fig. 3(a & c) are nearly identical to those in Fig. 3(d & b), respectively. They represent the same physical situation being computed using distinct modeling approaches and their agreement helps validate the accuracy of our modeling framework. It is instructive to re-plot the enhancement maps for the open and supported cells substituting the angle by the incident RF plane wave wavevector component in the xy-plane (kx or ky), as shown in Figure 5. Here the horizontal axis is the reduced wavevector XY component $k_{x,y}/2\pi = f\cos(\varphi,\theta)/c$, where $c$ is the speed of light and $f$ is the RF frequency. The supported cell with the period $p$ can couple the incident plane wave with an in-plane wavevector $(k_x, k_y)$ to a guided mode with wavevector $\left(k_x \pm \frac{2\pi}{p}m, k_y\right)$ with an integer $m$.

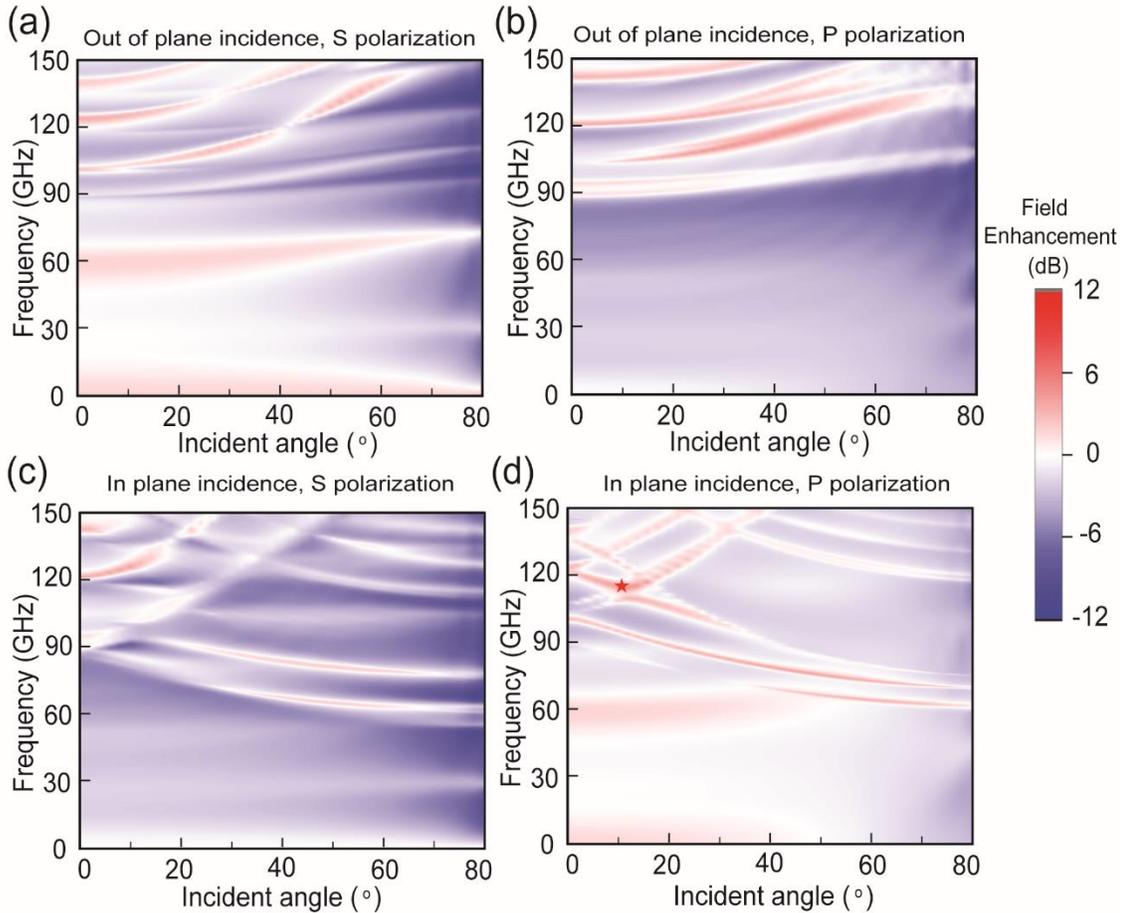

Fig. 4: Numerically calculated RF field enhancement inside a supported all-glass vapor cell with the variation of the angle of incidence for (a) Out-of-plane incidence, S polarization (b) Out-of-plane incidence, P polarization. (c) In-plane incidence, S polarization (b) In-plane incidence, P polarization. The amplitude of RF field enhancement indicated by the color bar. * Marks the location of highest observed enhancement.

In Fig. 5d, f, the in-plane incident wave $(k_x, 0)$ is coupled to modes with $\left(k_x \pm \frac{2\pi}{p}m, 0\right)$. Noting that the center of the horizontal axis corresponds to the edge of the first Brillouin zone for the guided modes in the periodic cell structure $\frac{\pi}{p} = 2\pi \cdot 0.25$ mm$^{-1}$, we observe the expected mirror symmetry in the enhancement with respect to this wavevector. We also see the straight-line dispersion relationships for the guided modes except near the various points of avoided crossings, and the folding near 0. In Fig. 5(c & e), the out-of-plane incident wave $(0, k_y)$ is coupled to modes with $\left(\pm\frac{2\pi}{p}m, k_y\right)$, and we observe the monotonically increasing frequencies as a function of $k_y$, with the fixed $k_x = \pm\frac{2\pi}{p}m$. For both planes of incidence some guided modes seem to be coupled from either incident polarization, while others appear to obey symmetry-based polarization selection rules. While we have described the underlying physics explaining the resonant features of the numerically observed enhancement, a detailed study of the specific guided modes supported by the cell and their dispersion relationships underlying the resonant enhancement is beyond the scope of the present study. Importantly, the sharp grating resonant enhancement modeled here results from the use of low loss dielectric materials. To illustrate this fully and to investigate the effect of material composition on RF field enhancement, we analyzed a supported cell configuration using a highly doped silicon instead of glass for the grating-structured middle layer, as depicted in the 3D schematic with blue coloration in Fig. 6(a), with all periodic unit cell dimensions identical to those of the structured cell in Fig. 1(c & d). The glass windows remain unchanged from the previous model, while for Si we have used the dielectric constant of 11.7 and the conductivity of 10 S/m resulting in the frequency dependent dielectric constant of $\varepsilon_{Si}(f) = \varepsilon' - i\varepsilon'' = 11.7 - i(\frac{10}{2\pi f \varepsilon_0})$. We have conducted identical numerical analysis of this Si supported cell, with the results presented in Fig. 6. While we observe a larger enhancement in the low frequency electrostatic limit for the ridge-normal fields (Fig.6 b, f, e & i) due to the larger dielectric constant of Si, the drastic difference is the field suppression under almost all other conditions as well as the lack of sharp enhancement features. The materials losses prevent the RF power buildup and field enhancement in the supported Si cell, other than the electrostatic enhancement observed at low frequencies.

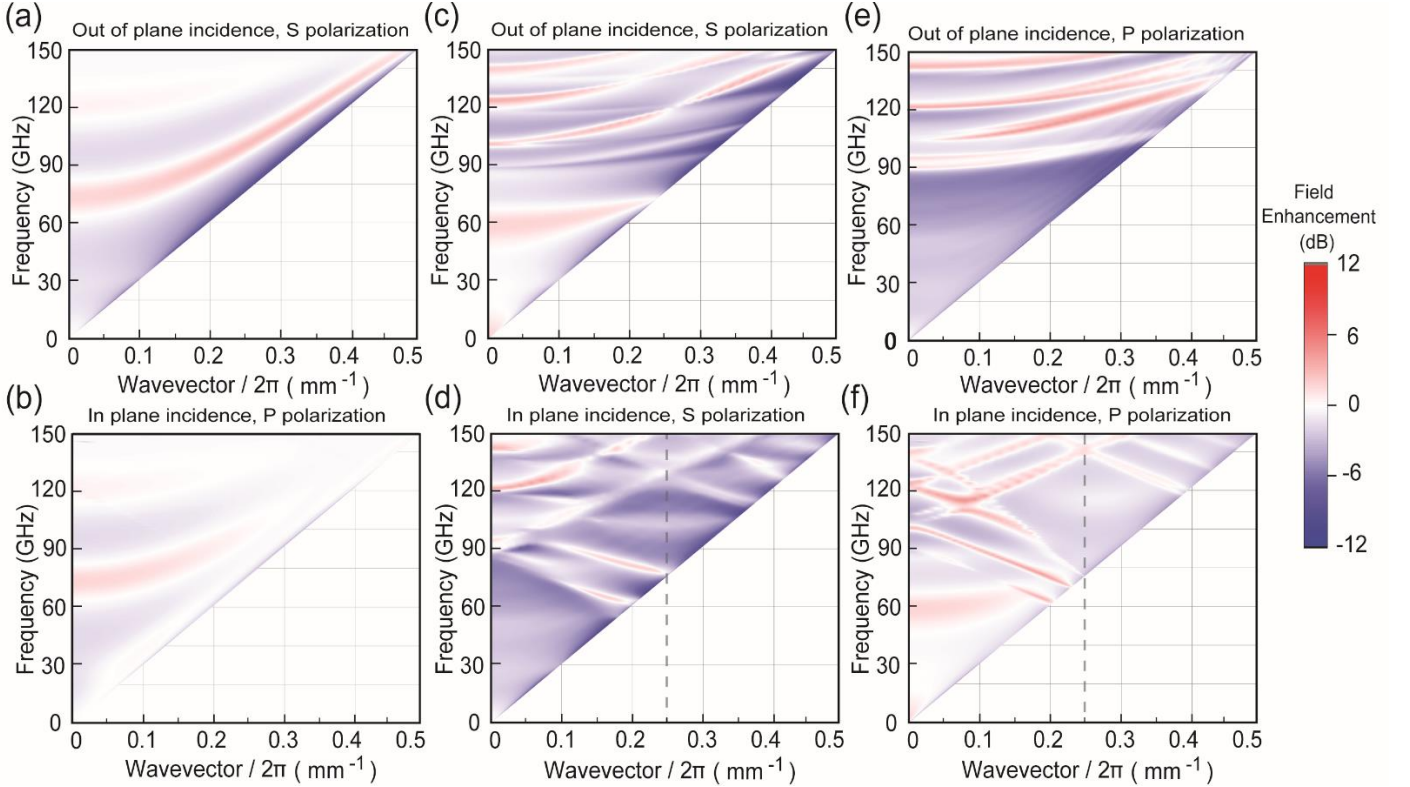

Fig. 5: Numerically calculated RF field enhancement versus incident plane wave frequency and wavevector component in the plane of the cell (xy). (a) and (b) correspond to the open vapor cell under S polarization and P polarization, respectively. (c)–(f) illustrate the supported vapor cell with out-of-plane incidence, S polarization (c), in-plane incidence, S polarization (d), out-of-plane incidence, P polarization (e), and in-plane incidence P polarization (f). The horizontal axis spans of $k_x$,y = 0.5 mm$^{-1}$ is equal to 1/p, where $p$ = 2 mm is the supported cell structure period. For (d) and (f) the $k_x$=0.25 mm$^{-1}$ corresponds to the edge of the first Brillouin zone (dashed line) for the RF bound modes in the periodic structure explaining the mirror symmetry of the enhancement w.r.t. that point.

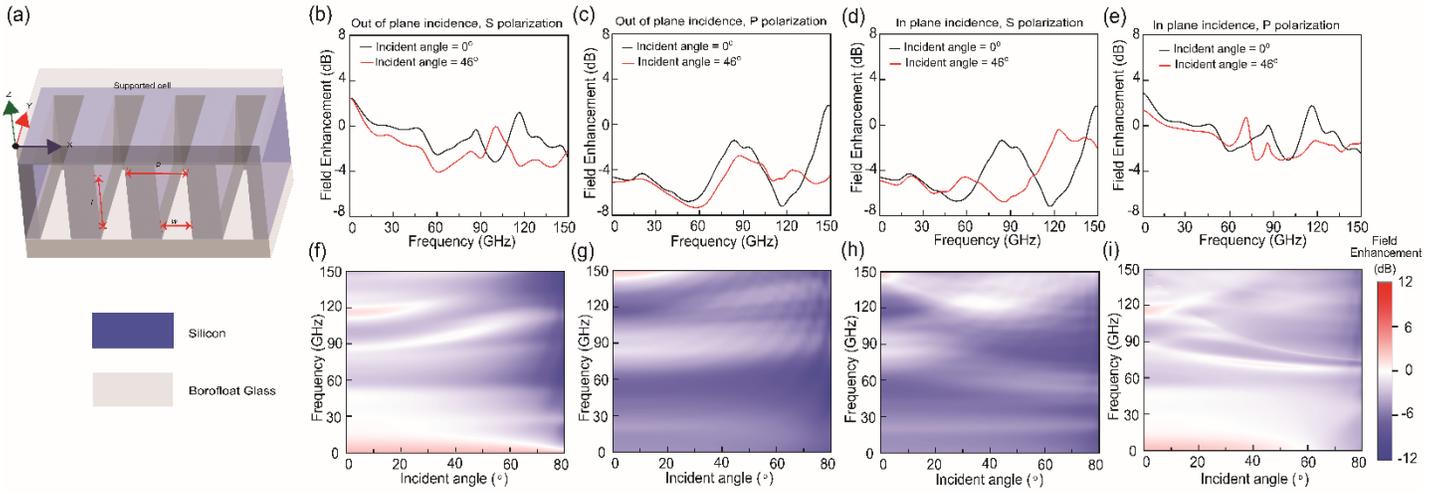

Fig. 6:(a) Illustrates a 3D schematic of the supported Si-glass vapor cell design, featuring a unit cell with a periodically structured, highly doped silicon grating (blue) centrally positioned between two parallel Borofloat glass (gray) layers. The structure dimensions are identical to the supported glass cell. (b–e) Show the numerically calculated RF field enhancement for two incident angles (0° and 46°), represented by black and red solid lines, respectively, under four excitation conditions: (b) out-of-plane incident S polarization, (c) out-of-plane incident P polarization, (d) in-plane incident S polarization, and (e) in-plane incident P polarization. (f–i) present the corresponding RF field enhancements as functions of frequency and angle of incidence for each polarization state.

III. DISCUSSION:

We have performed a comprehensive finite-element frequency domain electromagnetic modeling of millimeter waves interacting with miniaturized vapor cell geometries designed for Rydberg atom-based electrometry. We have analyzed the enhancement of the RF fields seen by the atomic vapor confined inside the cell cavities due to the cell structure and material, comparing a simple open cell geometry with periodically structured supported cells made out of either entirely of low loss dielectric glass or containing a lossy doped Si structured middle layer. In the open cell our simulations reveal the expected standing wave resonance field enhancement due to the reflections from the parallel glass windows. In the supported cell multiple sharp resonances with up to ≈2.9x field enhancements (>8x RF power density enhancement) are observed at the frequencies above approximately 60 GHz. These frequency and angle dependent enhancements are attributed to the grating resonances, i.e. the periodic structure of the cell coupling the incoming plane wave to guided modes inside the cell. This illustrates the novel engineering possibilities afforded by making cells entirely out of low loss dielectric materials, applying integrated photonic dielectric metamaterial concepts to the millimeter waves. At low frequencies, smaller but significant field enhancements within the cell are also shown, arising from the nearfields of the dielectric structures. The low loss nature of the dielectric is crucial to achieve resonant enhancements, which is entirely absent in cells made using the middle structures made of conductive lossy Si rather than glass. Angle- and k-vector-resolved field enhancement maps further illustrate the tunable and anisotropic nature of RF coupling in these microfabricated cells, with geometrical parameters and material selection offering new degrees of freedom for engineering spectrally and directionally selective electrometers. These results serve as a predictive framework for optimizing vapor cell designs to enhance field sensitivity and spectral selectivity, with direct applications in chip-scale Rydberg-based quantum sensors and RF imaging arrays. Future work will experimentally validate these results, extend this modeling approach to fully three-dimensional geometries and explore integration with on-chip photonic interfaces for laser excitation and detection. Such efforts will be pivotal in realizing scalable and portable platforms for precision electrometry using Rydberg atoms.

**References:**

[1] Sedlacek, Jonathon A., et al. "Microwave electrometry with Rydberg atoms in a vapour cell using bright atomic resonances." Nature physics 8.11 (2012): 819-824.

[2] Browaeys, Antoine, and Thierry Lahaye. "Many-body physics with individually controlled Rydberg atoms." Nature Physics 16.2 (2020): 132-142.

[3] Stebbings, R. F., and F. B. Dunning, eds. Rydberg states of atoms and molecules. Cambridge University Press, 1983.

[4] Schlossberger, N., Prajapati, N., Berweger, S. *et al.* Rydberg states of alkali atoms in atomic vapour as SI-traceable field probes and communications receivers. *Nat Rev Phys* **6**, 606–620 (2024).

[5] C. L. Holloway *et al*., "Broadband Rydberg Atom-Based Electric-Field Probe for SI-Traceable, Self-Calibrated Measurements," in *IEEE Transactions on Antennas and Propagation*, vol. 62, no. 12, pp. 6169-6182, Dec. 2014, doi: 10.1109/TAP.2014.2360208.

[6] Sedlacek, J_A, et al. "Atom-based vector microwave electrometry using rubidium Rydberg atoms in a vapor cell." Physical review letters 111.6 (2013): 063001.

[7] Miller, Stephanie A., David A. Anderson, and Georg Raithel. "Radio-frequency-modulated Rydberg states in a vapor cell." New Journal of Physics 18.5 (2016): 053017.

[8] Fan, Haoquan, et al. "Effect of vapor-cell geometry on Rydberg-atom-based measurements of radio-frequency electric fields." Physical Review Applied 4.4 (2015): 044015.

[9] Zhang, Linjie, et al. "Vapor cell geometry effect on Rydberg atom-based microwave electric field measurement." Chinese Physics B 27.3 (2018): 033201.

[10] Perez, Maximillian A., et al. "Rubidium vapor cell with integrated Bragg reflectors for compact atomic MEMS." Sensors and Actuators A: Physical 154.2 (2009): 295-303.

[11] Anderson, David A., et al. "Optical measurements of strong microwave fields with Rydberg atoms in a vapor cell." Physical Review Applied 5.3 (2016): 034003.

[12] Noaman, M., et al. "Vapor cell characterization and optimization for applications in Rydberg atom-based radio frequency sensing." Quantum Sensing, Imaging, and Precision Metrology. Vol. 12447. SPIE, 2023.

[13] Kübler, Harald, et al. "Coherent excitation of Rydberg atoms in micrometre-sized atomic vapour cells." Nature Photonics 4.2 (2010): 112-116.

[14] Wu, Bo, et al. "Dependence of Rydberg-atom-based sensor performance on different Rydberg atom populations in one atomic-vapor cell." Chinese Physics B 33.2 (2024): 024205.

[15] Amarloo, H., Noaman, M., Yu, SP. *et al.* A photonic crystal receiver for Rydberg atom-based sensing. *Commun Eng* **4**, 70 (2025).

[16] Artusio-Glimpse, Alexandra B., et al. "Wafer-level fabrication of fully-dielectric vapor cells for Rydberg atom electrometry." arXiv preprint arXiv:2503.15433 (2025).

[17] Rodriguez-Cano, Rocio, et al. "Novel Glass Material with Low Loss and Permittivity for 5G/6G Integrated Circuits." 2024 IEEE INC-USNC-URSI Radio Science Meeting (Joint with AP-S Symposium). IEEE, 2024.

[18] Any mention of commercial products is for information only; it does not imply recommendation or endorsement by NIST.